\documentclass[sigconf,nonacm]{acmart}

\usepackage{graphicx}

\renewcommand\footnotetextcopyrightpermission[1]{}
\begin{document}

\title{The Inference Bottleneck: Antitrust and Neutrality Duties in the Age of Cognitive Infrastructure}

\author{Gaston Besanson}
\affiliation{%
  \institution{Universidad Torcuato Di Tella}
  \city{Buenos Aires}
  \country{Argentina}
}

\author{Marcelo Celani}
\affiliation{%
  \institution{Universidad Torcuato Di Tella}
  \city{Buenos Aires}
  \country{Argentina}
}

\begin{abstract}
As generative AI commercializes, competitive advantage is shifting from one-time model training toward continuous inference, distribution, and routing. At the frontier, large-scale inference can function as cognitive infrastructure: a bottleneck input that downstream applications rely on to compete, controlled by firms that often compete downstream through integrated assistants, productivity suites, and developer tooling. Foreclosure risk is not limited to price. It can be executed through non-price discrimination (latency, throughput, error rates, context limits, feature gating) and, where models select tools and services, through steering and default routing that is difficult to observe and harder to litigate.

This essay makes three moves. First, it defines cognitive infrastructure as a falsifiable concept built around measurable reliance, vertical incentives, and discrimination capacity, without assuming a clean market definition. Second, it frames theories of harm using raising-rivals'-costs logic for vertically related and platform markets, where foreclosure can be profitable without anticompetitive pricing. Third, it proposes Neutral Inference: a targeted, auditable conduct approach built around (i) quality-of-service parity, (ii) routing transparency, and (iii) FRAND-style non-discrimination for similarly situated buyers, applied only when observable evidence indicates functional gatekeeper status.
\end{abstract}


\maketitle

\section{Introduction: From Models to Infrastructure Power}

Antitrust meets new technologies with an old question: when does control over a core input become the power to pick winners in vertically related markets? In the generative AI era, that question increasingly travels through inference and the interfaces that deliver it.

The commercial control point is not only who can train large models, but who can reliably serve frontier-grade inference at scale under technical and contractual terms downstream developers can depend on, and who can route users through AI interfaces that increasingly substitute for traditional distribution channels. Enforcement remarks and policy discussions have highlighted the risk pattern often described as ``open early / closed late''---generous early access that may induce dependence, followed by tighter access or degraded external quality once switching costs and distribution consolidation increase.~\cite{holyoak2025}

This essay's claim is narrow: if inference becomes a bottleneck input for a set of downstream applications, foreclosure, when it occurs, will often be implemented through non-price discrimination and steering, not only through posted prices.

\section{Why Inference Is the Bottleneck Surface}

Inference is the commercial surface through which providers differentiate and monetize: token-based pricing, throughput tiers, context limits, realtime endpoints, and capability wrappers (tool use, batch modes, enterprise governance). Public pricing and documentation show how access is materialized primarily as inference consumption and service features, not as model weights.~\cite{openai2026}

This packaging matters for competition because it creates multiple discrimination levers that are: (i) high impact (latency and reliability can determine viability for copilots, assistants, and real-time systems), (ii) low visibility (outsiders cannot easily observe serving priority, queue placement, or eligibility policies), and (iii) plausibly justified (differences can be attributed to capacity, safety, or abuse prevention).

OECD work in late 2025 emphasizes that competition risks are not confined to training and that AI adoption can reshape downstream competitive dynamics.~\cite{oecd2025a,oecd2025b} Market-definition ambiguity reinforces the case for focusing on measurable conduct: foundation models can be inputs across many sectors while also competing with downstream applications at the same time.~\cite{holyoak2025} The functional question is whether a provider can act as a discriminatory bottleneck for downstream rivals.

\section{Defining ``Cognitive Infrastructure'' as a Falsifiable Concept}

\textbf{Definition (Cognitive Infrastructure).} An inference service functions as cognitive infrastructure when all three conditions hold:
\begin{itemize}
    \item[(A)] \textbf{Material reliance with costly substitution.} Downstream products rely on the service's performance and governance envelope (reliability, latency, context, tool-use features, safety/compliance assurances) such that switching to available alternatives would require non-trivial reengineering and re-validation and/or would cause a measurable degradation in cost, quality, or compliance outcomes.
    \item[(B)] \textbf{Control with vertical incentives.} Access terms and technical quality are controlled by a firm (or small set of firms) that is vertically integrated into downstream complements competing with its API customers. The concern is ``referee-and-player'' incentives where upstream governance choices can predictably shift downstream rivalry.~\cite{holyoak2025}
    \item[(C)] \textbf{Discrimination capacity (non-price).} The controller can raise rivals' effective costs or reduce rivals' quality through non-price discrimination---QoS differentials, feature gating, model/version access, routing and eligibility policies---while maintaining facially neutral pricing.
\end{itemize}

\textbf{Testable predictions and minimal audit protocol.} (1) \emph{QoS wedge}: holding region, load, and contracted tier constant, first-party complements exhibit systematically better QoS (p50/p95/p99 latency, timeout/error rates, quota denials) than similarly situated third-party API users. (2) \emph{Routing bias}: where the model routes tasks to tools or services, routing outcomes favor the provider's own services or preferred partners at rates inconsistent with objective quality, coverage, or relevance benchmarks. A feasible audit uses repeated measurements over a sampling window; controls for region, tier, and request classes; and flags sustained deltas above a practical threshold (e.g., ${>}15\%$ in p95/p99 or denial rates over a rolling 7-day window) absent documented objective criteria. This does not require private user data or model-weight disclosure. Foreclosure concerns do not require strict indispensability in an essential-facilities sense; the question is whether conduct materially raises rivals' effective costs or degrades their ability to compete.~\cite{aplicacion2025,salop1983,salop1987}

\section{Theory of Harm and Mechanisms}

Foreclosure in inference markets is unlikely to appear as predatory pricing or explicit refusal to supply. It may appear as small, persistent degradations and steering that shift demand over time. Raising-rivals'-costs (RRC) theory provides a framework: a vertically integrated firm can profit by increasing rivals' effective costs (or degrading their quality) and recapturing demand, even without below-cost pricing.~\cite{salop1983,salop1987} Platform economics strengthens this logic in two-sided and ecosystem markets, where access rules and visibility/routing conditions can raise rivals' costs without changing a headline price.~\cite{armstrong2006,rochet2003} Related work on competing ecosystems predicts that total exclusion may be suboptimal; keeping rivals ``living but not thriving'' can preserve ecosystem breadth while relaxing competition against the gatekeeper's own downstream complements.~\cite{bisceglia2022} Self-preferencing can likewise be implemented through product design that degrades interoperability or selectively withholds advanced features.~\cite{motta2023}

Under cognitive-infrastructure conditions, three mechanisms matter: \emph{(i) QoS discrimination} (latency, throughput, error rates), where persistent QoS deltas reduce rival quality and force costlier mitigations; \emph{(ii) feature gating} (context, tool use, enterprise wrappers), where capabilities or timing advantages are withheld from similarly situated third parties; and \emph{(iii) steering and default routing} in agentic flows, where opaque selection reduces rivals' effective access to demand absent auditable routing artifacts. First-party latency advantages can also reflect legitimate architectural efficiency (colocation, shared caches, fewer hops). The audit protocol must therefore distinguish architectural efficiency from artificial degradation by controlling for region/topology and focusing on residual, persistent wedges absent objective criteria.

\section{Neutral Inference: Targeted, Auditable Neutrality Duties}

Neutral Inference is a conduct approach designed to preserve contestability in downstream markets by constraining discrimination at a bottleneck input only when observable evidence indicates functional gatekeeper status.

\textbf{Institutional fit.} U.S.\ antitrust enforcers are ``enforcers, not regulators,'' and courts are reluctant to impose broad ex ante access mandates.~\cite{holyoak2025} Trinko reinforces limits around compelled dealing as a general remedy.~\cite{trinko2004} The EU has adopted contestability and fairness duties for gatekeepers under the DMA, including constraints relevant to self-preferencing and steering.~\cite{dma2022,ec2024dma} The AI Act and the GPAI Code of Practice create transparency and risk-management artifacts that can support technical auditing without requiring disclosure of proprietary weights.~\cite{aiact2024,gpai2025}

\textbf{Positioning.} Neutral Inference is best read as an operational design pattern: it translates non-discrimination and anti-steering concerns into auditable technical artifacts (QoS parity tests and routing logs). Routing-log transparency concerns selection outcomes and eligibility constraints, not chain-of-thought disclosure.

\textbf{Designation triggers.} Dependency and switching costs across a meaningful set of downstream products; sustained performance or governance gaps versus close substitutes on metrics customers buy; and vertical expansion into distribution through major downstream complements that compete with API customers and shape routing.~\cite{holyoak2025,oecd2025a,oecd2025b}

\textbf{Three obligations.} (1) \emph{QoS parity}: no materially better inference QoS for first-party complements than for similarly situated third parties, after controlling for objective cost and risk drivers, with standardized aggregated disclosures. (2) \emph{Routing transparency}: auditable routing records capturing eligible tool sets, selected tools (and rank where applicable), and whether commercial relationships constrained eligibility or influenced ranking. (3) \emph{FRAND-style non-discrimination}: non-discriminatory access terms for similarly situated buyers, with fast dispute resolution. A workable enforcement model resembles compliance auditing more than decade-long litigation: periodic audits; standardized test suites for QoS parity and routing bias; interim measures when discrimination is detected; and penalties tied to inference revenue.

\begin{figure}[t]
    \centering
    \includegraphics[width=\columnwidth]{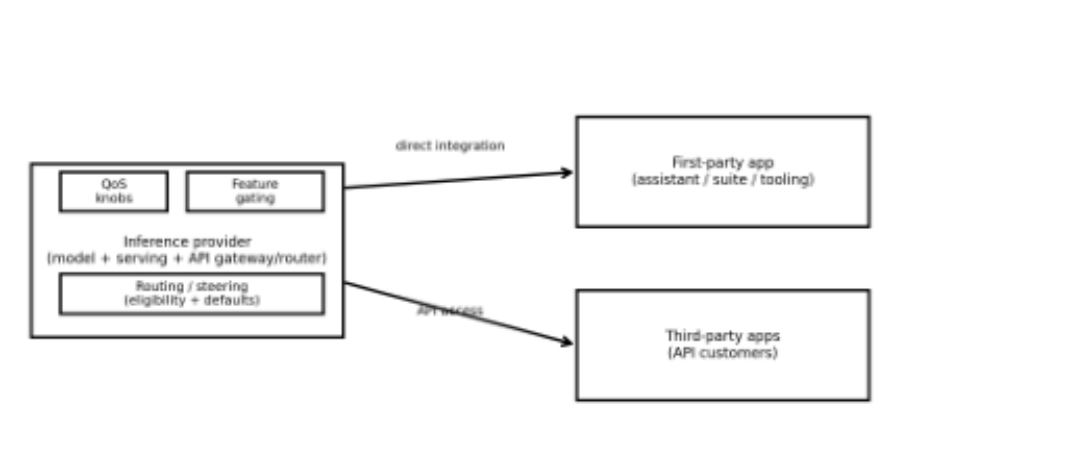}
    \caption{Inference bottleneck architecture. A provider controlling model serving and an API gateway/router can implement non-price discrimination and steering through (i) QoS knobs (queue priority, rate limits), (ii) feature gating (wrappers/version access), and (iii) routing/steering (tool eligibility/defaults), without changing headline API prices.}
    \label{fig:architecture}
\end{figure}

\section{Conclusion}

The central competitive risk in the next phase of AI is not only concentration in training. It is the inference bottleneck: scalable cognition integrated into downstream products and mediated through non-price discrimination and opaque routing. Neutral Inference offers a narrow response---QoS parity, routing transparency, and FRAND-style non-discrimination---applied only when observable evidence indicates functional gatekeeper status. The objective is not to punish scale. It is to preserve contestability before the infrastructure layer quietly acquires the power to determine which applications can exist.

\bibliographystyle{ACM-Reference-Format}

\begin{thebibliography}{18}
\bibitem{holyoak2025c}
Holyoak, supra note 1 (FTC/DOJ as enforcers, not regulators; reluctance to ex ante rules).
\end{thebibliography}

\end{document}